\documentclass[prb, showpacs,twocolumn, preprintnumbers]{revtex4}
\usepackage {graphicx}

\begin{document}

\title{Enhancement of ferromagnetism upon thermal annealing in pure ZnO}

\author{S. Banerjee\footnote{Email:sangambanerjee@saha.ac.in}}
\address{Surface Physics Division, Saha Insitute of Nuclear Physics, 1/AF Bidhannagar, Kolkata 700 064, India}

\author{M. Mandal}
\address{Chemical Sciences Division, Saha Insitute of Nuclear Physics, 1/AF Bidhannagar, Kolkata 700 064, India}

\author{N. Gayathri\footnote{Present address: Material Science Section, Variable Energy Cyclotron Center, 1/AF Bidhannagar, Kolkata 700 064, India} and M.~Sardar}
\address{Material Science Division, Indira Gandhi Center for Atomic Research, Kalpakkam 603 102, India}

\begin{abstract}
We report here enhancement of ferromagnetism in pure ZnO upon thermal annealing with the ferromagnetic transition temperature T$_c$ above room temperature. We observe a finite coercive field upto 300K and a finite thermoremanent magnetization upto 340K for the annealed sample. We propose that magnetic moments can form at anionic vacancy clusters. Ferromagnetism can occur due to either superexchange between vacancy clusters via isolated F$^+$ centers, or through a limited electron delocalization between vacancy clusters. Isolated vacancy clusters or isolated F$^+$ centers give rise to a strong paramagnetic like behaviour below 10K. 
\end{abstract}

\maketitle

Recently ferromagnetism (FM) has been observed in undoped wide band gap semiconductor such as TiO$_2$, ZnO, HfO$_2$, SnO$_2$, In$_2$O$_3$, Al$_2$O$_3$, CeO$_2$ etc. \cite{VenkatesanNature,CoeyPRB,HongPRB2006,Yoon,HongAPL2006,Schwartz2004,Radovanovic,HongPRB2005,CNR,HongJPC2007} in reduced dimensional form such as nanoparticle and thin films. Observation of ferromagnetism in these undoped systems created more excitement and threw open a wider debate as to the origin of magnetism in these wide band gap semiconductors. The ferromagnetism in these low dimensional systems has been attributed to oxygen defects and has been suggested as arising due to low dimensionality \cite{CNR}. The saturation moments in all these materials are typically $10^{-4}-10^{-3}$ emu/gm. But, till todate, there has not been any explanation of how exactly the ferromagnetism arises due to the oxygen defects. The nature and specific organisation of the defects which can give rise to ferromagnetism is an important issue. There exists evidences that, perfect crystallinity and having more oxygen can degrade FM \cite{Gamelin}. Recently it has been shown that oxygen vacancy occuring in the O(3) sites in bulk Ga$_2$O$_3$ can give rise to ferromagnetism \cite{Sreedharan}.  Another observation recently reported is that ZnO grown by sol-gel technique also shows ferromagnetic behaviour \cite{condmatMn}. The most interesting observation reported in the same paper was that the ferromagnetism of the ZnO was destroyed upon intentionally doping with magnetic impurity such as Mn \cite{condmatMn}. In this letter, we report on the observation of ferromagnetism in pure ZnO grown by micellar method. Transition metal doped ZnO is being commonly used for magneto-optic applications but our experimental results opens up a different avenue of using pure ZnO itself (ie., without transition metal doping) for magneto-optic application. We also give a possible mechanism that we conjecture for the appearence of ferromagnetism in ZnO in this letter. 

ZnO has a simple wurtzite structure where all oxygens and Zn  sites are equivalent. We selected pure ZnO for our investigation for magnetization measurement. The samples were prepared by miceller method using ZnSO$_4$ (1 m mol) and Sodium dodecyl sulphate (SDS) micelles (100 ml of 10$^{-2}$ M). After thorough stirring and boiling 1 m mol NaOH in 1 ml aqueous solution was added slowly. Ethyl alcohol was added after refluxing the mixture for 1 hour and allowed to stand overnight. The precipitated ZnO white powder obtained after centrifuging was heated at 50$^o$C for 24 hours to dry. This as-prepared sample was annealed at 900$^o$C for 2 hours and the resultant powder was faint yellow in colour. Structural characterization and chemical purity of the sample was carried out using XRD, SEM, EDAX and photoluminescence \cite{condmatbulkzno}. The particle size was between 200 to 500 nm. The magnetic property was measured using MPMS-7 (Quantum Design). For the zero field cooled (ZFC) data, the sample was cooled down to 2K in the absence of magnetic field and the data was taken while warming up upto 340K in the presence of 100 Oe field. The field cooled data (FC) was taken while warming the sample after cooling it to 2K in the presence of the 100 Oe field. The hysteresis data were taken at various temperatures upto 300K and upto a field of 1 Tesla. 

\begin{figure}
\includegraphics[width=6cm,angle=270]{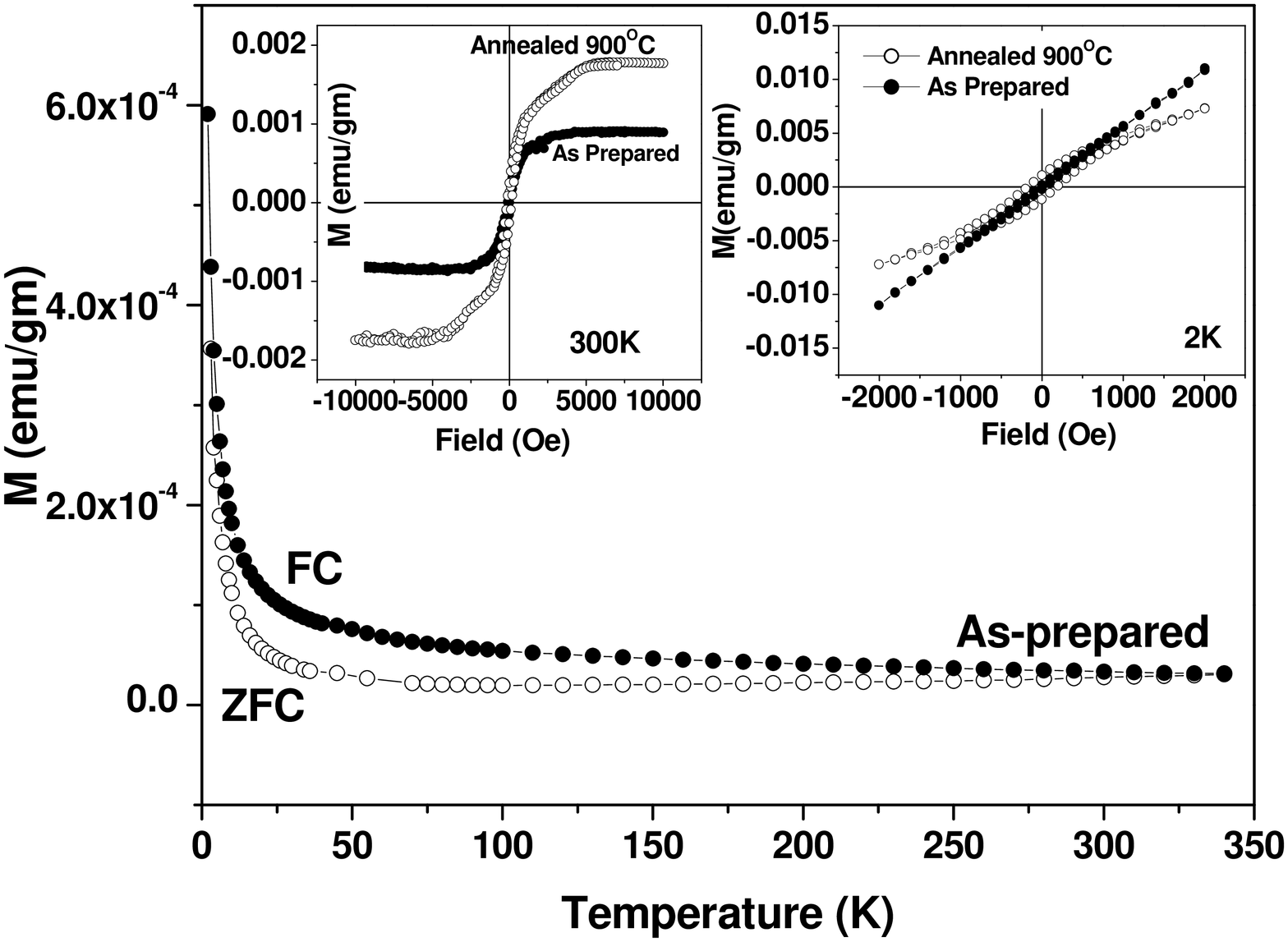}
\caption{ZFC and FC magnetisation curves of the as-prepared sample taken at 100 Oe field showing the distinct bifurcation around 340K. Insets: Comparison of M-H curves of the as-prepared and annealed (900${^O}$C) sample at 300K and 2K.}
\end{figure}

Fig. ~1 shows the ZFC and FC magnetisation curve taken at 100 Oe for the as-prepared sample. The distinct bifurcation of the FC and ZFC starts around 340K which is also seen from the finite thermoremanent magnetisation in the inset of fig.~2. Both the ZFC and the FC data show paramagnetic-like behaviour. The insets of fig.~1 shows the M-H data of the as-prepared sample taken at 2K upto 2000 Oe and at 300K upto 1 Tesla after the necessary background diamagnetic subtraction. There is no clear hystersis at both the temperatures. The 300K data shows S-shaped behaviour with saturation similar to a superparamagnetic behaviour. There is no saturation of magnetisation at 2K upto 2000 Oe. The M-H data in the insets have been compared with that of the annealed sample which will be further discussed below.

\begin{figure}
\includegraphics[width=6cm,angle=270]{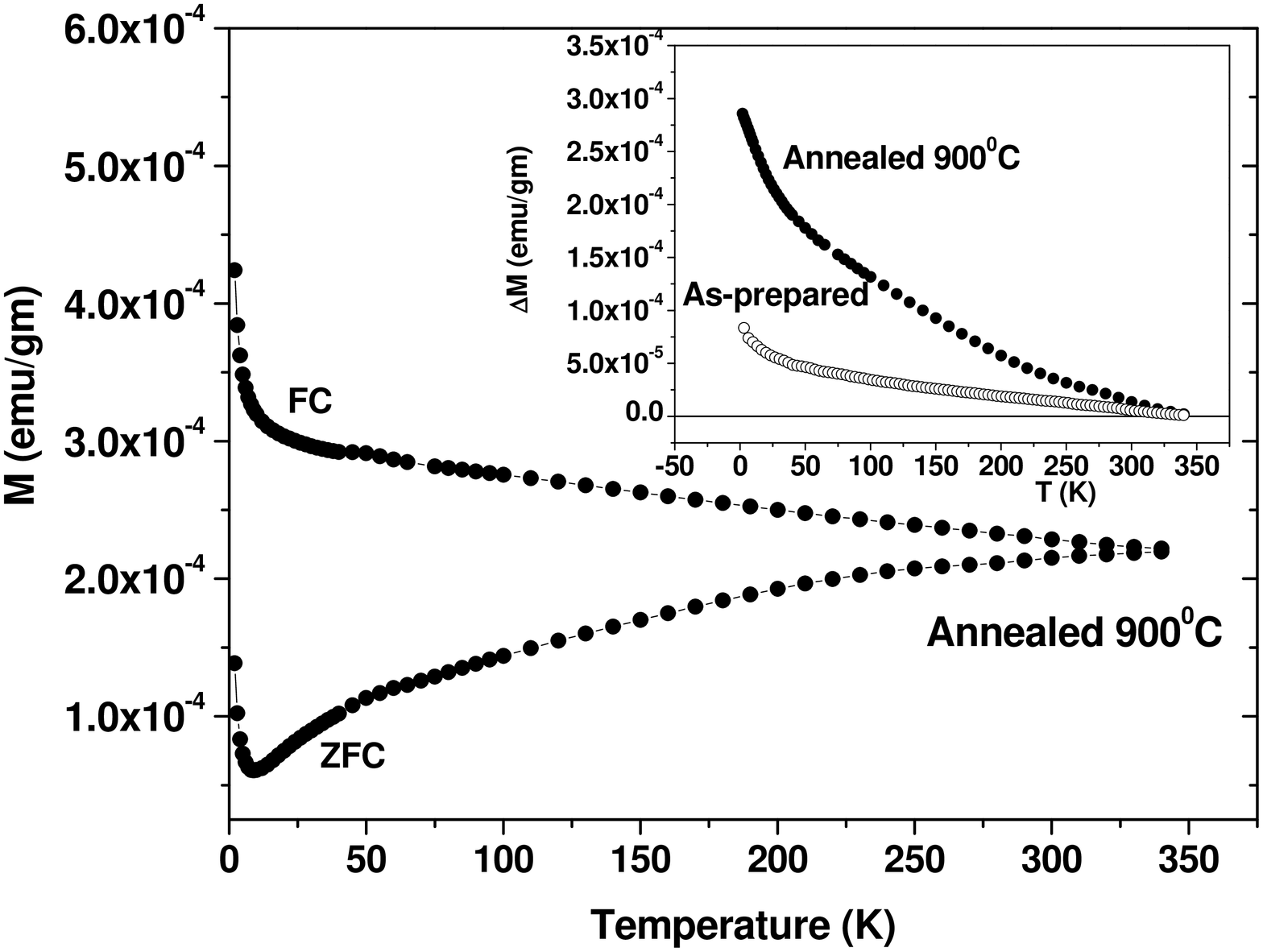}
\caption{ZFC and FC magnetisation curves of the 900$^o$C annealed sample taken at 100 Oe field showing the distinct bifurcation above 340K. Inset: The thermoremanent magnetisation of as-prepared and the annealed samples showing the finite value upto 340K}
\end{figure}

Fig. ~2 shows the ZFC and FC magnetisation curve taken at 100 Oe for the annealed sample. The distinct bifurcation of the FC and ZFC starts well above 340K. It is obvious from the inset that the thermoremanent magnetisation has enhanced significantly upon thermal annealing. The ZFC magnetisation increases as a function of temperature for T $>$ 10K indicating that there are blocked moments which start contributing to the magnetisation when the temperature is increased. The FC magnetisation decreases as the temperature increases, since the frozen moments start randomizing due to thermal energy. This behaviour is seen in most DMS systems. It is interesting to note that there is a sharp increase in the magnetisation in both the FC and the ZFC data below 10K. The origin of this strong paramagnetic behaviour will be discussed below.

\begin{figure}
\includegraphics[width=6cm,angle=270]{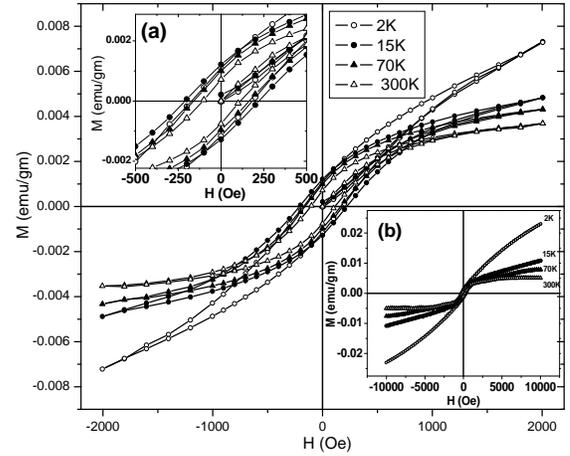}
\caption{M-H loops of the 900$^o$C annealed sample at T = 2K, 15K, 70K and 300K upto 2000 Oe showing the non-saturation. Inset(a): Expanded low field region of the M-H loops showing the finite coersive field. Inset(b): M-H loops at the same temperatures upto 1 Tesla field showing saturation atleast at 300K.}
\end{figure}

In fig.~3 we show the (M vs. H) hysteresis curves of the sample at 4 different temperatures upto 2000~Oe after the necessary background diamagnetic subtraction. We observe clear hysteresis at all temperatures upto 300K as seen in inset(a) of fig.~3. The inset(b) shows the hystersis upto 1 Tesla showing non-saturation of the magnetisation at 2K. Magnetisation saturation can be observed for 300K.

The comparision of the M-H data of the annealed and the as-prepared sample as shown in the inset of fig.~1 indicates that there is an enhancement in the ferromagnetic contribution on annealing. There is a higher coercivity for the annealed sample compared to the as-prepared sample at 2K. The saturation magnetization at 300K is also higher for the annealed sample indicating that there is definitely an increase in the ferromagnetic contribution on annealing. This comparison between the as-prepared and the annealed sample clearly indicates that the observed enhancement of ferromagnetism is not due to any magnetic impurity in the sample rather it is an annealing effect.
This phenomena of enhancement of ferromagnetism upon thermal annealing is explained now.

One way in which magnetic moments could arise from atomic defects in ZnO is as follows. In case of single neutral oxygen vacancy the electronic states at the O vacancy can be obtained using molecular orbital (MO) theory. At each of the four Zn ions next to an oxygen vacancy, one sp$^3$ hybrid of 4s and 4p orbitals can be formed which points towards the vacancy. From these four orbitals one a$_1$ and three t$_2$ (degenerate) orbitals centered at the O vacancy are formed. Careful local cluster calculations, by many groups gives a splitting of 2.7 eV or more, between $a_1$ and t$_2$ levels \cite{Fink}. We do a simple tight binding modelling with 4 orbitals, with an inter-orbital hopping matrix element of $t=-0.7$ eV and get the same energy level structure as obtained from other more refined calculations \cite{Fink}. So a neutral oxygen vacancy will have 2 electrons in a$_1$ state, i.e a singlet with spin zero because this splitting is greater than the Hund's coupling energy of $\sim$  1eV. Hence a single neutral oxygen vacancy will not have any net magnetic moment. Extending this calculation to vacancy clusters we have found that for a cluster with more than three oxygen vacancies, the seperation between the energy levels are lower than the Hund's coupling energy thus giving rise to net moment in the cluster. On annealing oxygen vacancies can migrate and form clusters because it reduces the bond energies favouring decrease in the strain field energy \cite{THOMPSON}. Such oxygen vacancy clusters have also been observed in HTSC oxides \cite{YBCO} upon annealing which shows up as the peak effect in the current density measurements.

\begin{figure}
\includegraphics[width=6cm,angle=270]{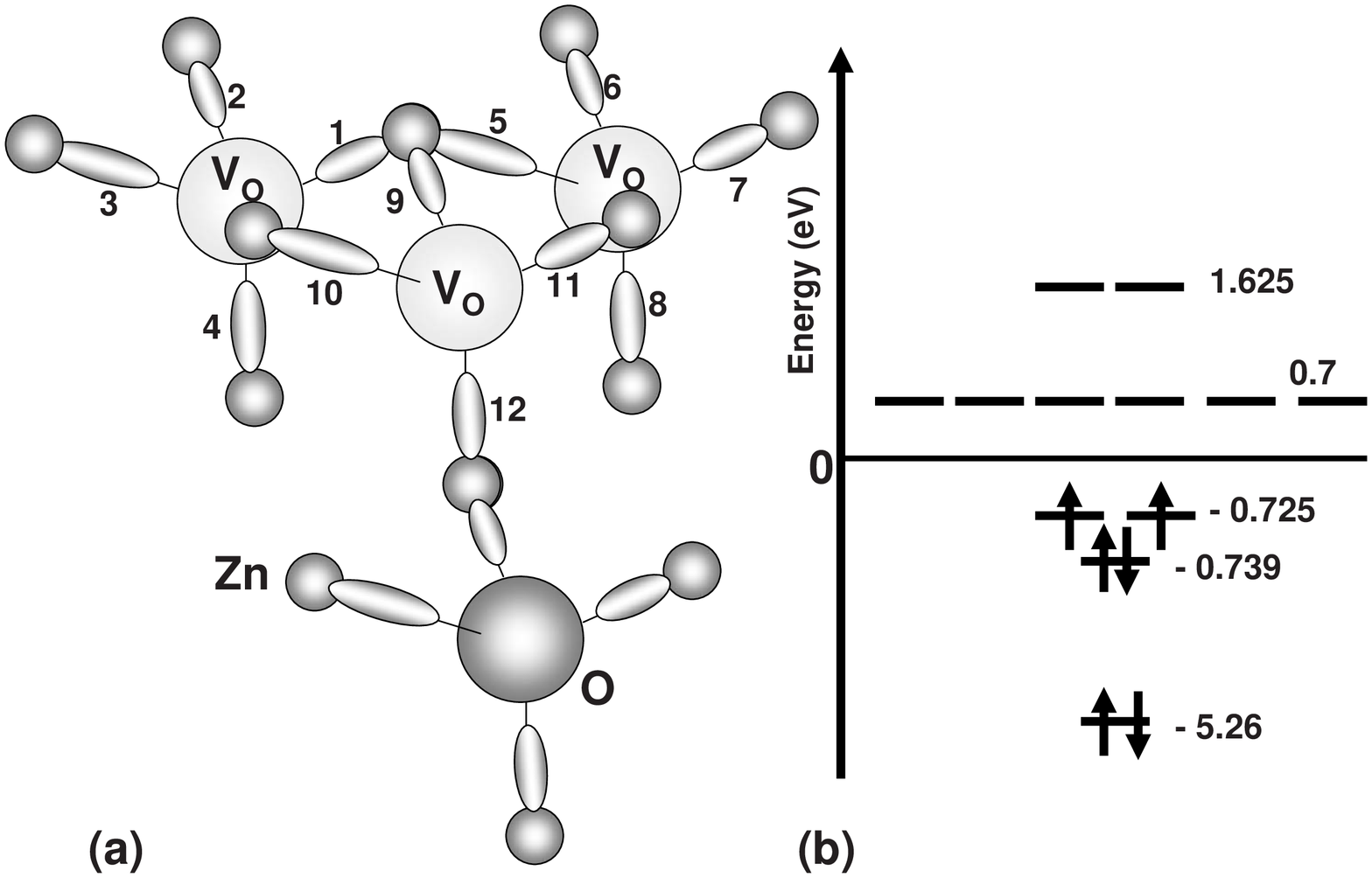}
\caption{(a) Tetrahedral structure of ZnO showing three oxygen vacancies labeled as V$_O$ and the orbitals (labled 1 to 12) used for calculating the energy levels for the three oxygen vacancy cluster. (b) The calculated energy levels showing the occupancy of 6 electrons (each neutral oxygen vacancy has an occupancy of 2 electrons) and hence a net moment arising due to degeneracy of the third and the fourth energy levels (-0.725 eV).}
\end{figure}

Here, we will consider a cluster with three near neighbour oxygen vacancy configuration, with 12 basis states around 10 Zn atoms (of which one Zn atom is shared between 3 oxygen vacancies, which are in the basal plane $\perp$ to the z axis) as shown in fig. 4(a). We assume that the hopping matrix element between the orbitals (three in number) centered about the Zn atom, which is shared by three vacancies larger by a factor of two, ie., $2t$, which is reasonable since these orbitals are centered about the same Zn atom, and assuming a small matrix element of -0.2 eV between Zn orbitals surrounding one O vacancy and the Zn orbitals surrounding the neighbouring O vacancy,
we get energy levels at, -5.26, -0.739, -0.725 (twofold degenerate), 0.7 (sixfold degenerate), 1.625(twofold degenerate) and is shown schematically in fig. 4(b). We see that with a small cluster of three neutral oxygen vacancy containing six electrons in total, we can easily generate a spin one configuration. This is certainly not an exact calculation, but we point it out to show, that even if one cannot have a moment forming at isolated oxygen/Zn vacancy site, yet there is a strong possibility of forming net moment for a cluster of vacancies. Our choice of hopping matrix elements for three vacancy problem is such that the third and the fourth energy levels are degenerate (-0.725 eV), but it doesnot have to be so restrictive. So long as the level separation between these two levels are less than the effective Hund's coupling between the molecular orbitals, a net spin can develop. Such isolated vacancy clusters with no interaction between them can only give paramagnetism, which we really see at the lowest temperatures (below T=$10$ K) where moments starts building up in our magnetisation data, as well as the non-saturation of net moment in the M-H data at T=$2$ K. To get a weak ferromagnet with clear hysteresis (as is observed) one has to go beyond this, and we can only speculate some reasons at this stage of our work:

(1) Isolated F$^+$ centers (oxygen vacancy with one electron) which were seen in EPR experiments \cite{Pan} with spin $1\over 2$, can couple antiferromagnetically with vacancy clusters  (superexchange) to give a ferrimagnet with a net moment. 

(2) There could be limited delocalisation of electrons between unequal size vacancy clusters giving rise to ferromagnetism (double exchange).

The various ways in which vacancy clusters can give rise to a ferromagetic state, needs to be explored both experimentally and theoretically. To get an estimate of the order of magnitude of moments that can arise from oxygen vacancies, we assume the oxygen nonstoichiometry to be $0.1 \%$ (i.e one in 1000 oxygen atom is absent). If only one tenth of these vacancies generate a moment of $1 \mu_B$, then we get a moment of $4 \times 10^{-3}$ emu/gm which is close to the saturation moment at T=15 K and 2000 ~Oe seen by us.

To summarise, we have reported here observation of enhancement of ferromagnetic contribution in pure ZnO sample upon thermal annealing. The as-prepared sample shows very weak ferromagnetic contribution compared to the annealed sample. The enhancement of ferromagnetism has been attributed to formation of oxygen vacancy clusters upon thermal annealing. The finite thermoremanent magnetization obtained from the difference between the ZFC and FC data upto 340K indicates that the ferromagnetic transition temperature T$_c$ of the pure ZnO sample investigated to be above 340K. We argue that large moments can be generated, if one has vacancy clusters of three or more oxygen vacancies. This observation indicates that if the already existing wide-band gap metal oxides like ZnO used in optoelectronic devices can be made magnetic (without transition metal doping) due to small oxygen vacancy cluster then a different avenue of magneto-optic applications will be opened up.

{\bf Acknowledgement}: The authors would like to thank Prof. S. Chaudhury, IACS for the PL measurements, Prof. A. De for discussion, Mr. Souvik Banerjee for SEM and EDAX measurements and Mr. Sushanta Banerjee for XRD measurements.


\newpage

\begin{thebibliography}{99}
\bibitem{VenkatesanNature} M. Venkatesan, C. B. Fitzgerald and J. M. D. Coey, Nature {\bf 430} 630 (2004)
\bibitem{CoeyPRB} J. M. D. Coey, M. Venkatesan , P. Stamenov, C. B. Fitzgerald and L. S. Dorneles, Phys. Rev. B {\bf 72} 24450 (2005)
\bibitem{HongPRB2006}  N. H. Hong, J. Sakai, N. Poirot and V. Brize, Phys. Rev. B {\bf 73} 132404 (2006)
\bibitem{Yoon} S. D. Yoon, Y. Chen, A. Yang, T. L. Goodrich, X. Zou, D. A. Arena, K. Ziemer, C. vittoria and V. G. Harris, J. Phys: Condens. Matter {\bf 18} L355 (2006)
\bibitem{HongAPL2006} N. Hoa Hong, N. Poirot and J. Sakai, Appl. Phys. Lett. {\bf89} 042503 (2006)
\bibitem{Schwartz2004} D. A. Schwartz, and D. R. Gamelin, Adv. Mater. {\bf16} 2115 (2004)
\bibitem{Radovanovic} P. V. Radovanovic and D. R. Gamelin, Phys. Rev. Lett. {\bf91} 157202 (2003)
\bibitem{HongPRB2005} N. Hoa Hong, J. Sakai, N. T. Huong, N. Poirot and A. Ruyter, Phys. Rev. B {\bf72} 45336 (2005)
\bibitem{CNR} A. Sundaresan, R. Bhargavi, N. Rangarajan, U. Siddesh and C.N.R. Rao, Phys. Rev. B. {\bf74} 161306R (2006)
\bibitem{HongJPC2007} N. H. Hong, J. Sakai and V. Brize, J. Phys: Condens. Matter {\bf19} 036219(2007)
\bibitem{Gamelin} P. V. Radovanovic and D. R. Gamelin, Phys. Rev. Lett. {bf91}, 157202(2003), K. R. Kittilstved, N. S. Norberg and D. R. Gamelin, Phys. Rev. Lett. {\bf94}, 147209(2005), N. H. Hong, J. Sakai, N. T. Huong, N. Poirot and A. Ruyter, Phys. Rev. B{\bf72}, 045336(2005).
\bibitem{Sreedharan} V. Sridharan, S. Banerjee, M. Sardar, S. Dhara, N. Gayathri and V. S. Sastry, cond-mat/0701232
\bibitem{condmatMn} S. Banerjee, K. Rajendran, N.Gayathri, M. Sardar, S.Senthilkumar and V.Sengodan, cond-mat/0704.3541
\bibitem{condmatbulkzno} S. Banerjee, M. Mandal, N. Gayathri and M. Sardar, cond-mat/0702486v1
\bibitem{Fink} Karen Fink, Phys. Chem, Chem. Phys, {\bf 7},2999(2005), Vom Fachbereich, cond-mat/981134432.
\bibitem{THOMPSON} M.W. Thompson, Defects and Radiation Damage in metals, Cambridge University Press, Chapter 2 (1969) 
\bibitem{YBCO} H. Kupfer, Th. Wolf, C. Lessing, A. A. Zhukov, X. Lancon, R. Meier-Hirmer, W. Schauer, and H. Wuhl, Phys. Rev B, {\bf58}, 2886 (1998)
\bibitem{Pan} Dengyu Pan, Guoliang Xu, Liya Lv, Yuan Yong, Xiuwei Wang, Jianguo Wan, 
Guanghou Wang and Yunxia Sui, Appl. Phys. Letters, {\bf 89}, 082510(2006).
\end{thebibliography}
\end{document}